\documentclass[hyper]{JHEP} 

\usepackage{epsfig}

\newcommand\fverb{\setbox\pippobox=\hbox\bgroup\verb}

\newcommand\fverbdo{\egroup\medskip\noindent%

            \fbox{\unhbox\pippobox}\ }

\newcommand\fverbit{\egroup\item[\fbox{\unhbox\pippobox}]}

\newbox\pippobox

\title{Canonical Analysis of Non-Relativistic Particle
    and Superparticle}
\author{J. Kluso\v{n}\\
Department of
Theoretical Physics and Astrophysics\\
Faculty of Science, Masaryk University\\
Kotl\'{a}\v{r}sk\'{a} 2, 611 37, Brno\\
Czech Republic\\
E-mail: \email{klu@physics.muni.cz}} \preprint{}

 \abstract{We perform canonical analysis of non-relativistic
 particle in Newton-Cartan Background. Then we extend this analysis
to the case of non-relativistic superparticle in the same background.
We determine constraints structure of this theory and find generator
of $\kappa-$symmetry.}

\def\hpi{\hat{\pi}}

\def\bV{\mathbf{V}}
\def\bW{\mathbf{W}}

\def\be{\begin{equation}}

\def\ee{\end{equation}}

\def\bea{\begin{eqnarray}}

\def\eea{\end{eqnarray}}

\def\mH{\mathcal{H}}

\def\bM{\mathbf{M}}

\newcommand{\mG}{\mathcal{G}}

\newcommand{\bT}{\mathbf{T}}

\def\pb #1{\left\{#1\right\}}

\begin{document}
\section{Introduction and Summary}
Holography is very useful for the analysis of properties of strongly coupled
quantum field theories. Recently these ideas were extended to non-relativistic
theories since today it is well known that   non-relativistic holography  is very useful tool for the study of strongly correlated systems in condensed matter, for recent review see \cite{Hartnoll:2016apf}. Non-relativistic symmetries also have fundamental meaning in the recent proposal of renormalizable quantum theory of gravity known today as Ho\v{r}ava-Lifshitz gravity \cite{Horava:2009uw}, for recent review and extensive list of references, see \cite{Wang:2017brl}. There is also an interesting connection between Ho\v{r}ava-Lifshitz gravity
and Newton-Cartan gravity \cite{Hartong:2016yrf,Hartong:2015zia}. Newton-Cartan gravity is covariant and geometric reformulation of Newton gravity that is now very intensively studied, see for example
\cite{Bergshoeff:2017dqq,Bergshoeff:2017btm,Bergshoeff:2016lwr,
    Hartong:2015xda,Bergshoeff:2015uaa,Andringa:2010it,Bergshoeff:2015ija,Bergshoeff:2014uea} 
\footnote{See also 
	\cite{Banerjee:2014pya,Banerjee:2014nja,Banerjee:2015rca,Banerjee:2016laq,Banerjee:2016bbm}.}

Concept of non-relativistic physics also  emerged
in string theory  when strings and branes were analyzed at special backgrounds. At these special points of string moduli spaces  non-relativistic symmetries emerge in natural way
 \cite{Gomis:2000bd,Danielsson:2000gi} . These actions were obtained by non-relativistic "stringy" limit where time direction and one spatial direction along the string are large. The stringy limit of superstring in $AdS_5\times S^5$ was also formulated in
\cite{Gomis:2005pg} and it was argued here that it provides another soluble sector of AdS/CFT correspondence, for related work, see \cite{Sakaguchi:2006pg,Sakaguchi:2007zsa}.  Non-relativistic limit was further extended to the case of higher dimensional objects in string theory, as for example p-branes
\cite{Gomis:2004pw,Kluson:2006xi,Brugues:2004an,Gomis:2005bj}\footnote{For recent works, see
        \cite{Batlle:2016iel,Gomis:2016zur,Batlle:2017cfa,Kluson:2017djw,Kluson:2017ufb,Kluson:2017vwp}.}.

It is important to stress that these constructions of non-relativistic
objects are based on manifest separation of directions along which  the non-relativistic limit is taken and directions that are transverse to them. The first step for the more covariant formulation which corresponds to the particle in Newton-Cartan background was performed in \cite{Bergshoeff:2014gja} and further elaborated in \cite{Bergshoeff:2015uaa}. The structure of this action is very interesting and certainly deserves further study. In particular, it would be very useful to find Hamiltonian for this particle.
Our goal in the first part of this article
is to find the Hamiltonian formulation of the particle in Newton-Cartan and in Newton-Cartan-Hooke background. It turns out that this is non-trivial task
even in the bosonic case due to the complicated structure of the action. On the other hand when we determine Hamiltonian constraint we find that the canonical structure of this theory is trivial due to the fact that there is only one scalar first class constraint. A more interesting situation occurs when
we consider supersymmetric generalization of the non-relativistic particle.
As the first case we study Galilean  superparticle whose action was proposed in   \cite{Bergshoeff:2014gja}. We find its Hamiltonian form and identify
primary constraints. We show that fermionic constraints are the second class constraints that can be solved for the momenta conjugate to fermionic variables when we also obtain non-trivial Dirac brackets between fermionic variables.
Next we consider more interesting case corresponding to the $\kappa-$symmetric
non-relativistic particle action \cite{Bergshoeff:2014gja}. We again determine all primary constraints. Then the requirement of the preservation of the fermionic primary constraints determines corresponding Lagrange multipliers. In fact we find a linear combination of the fermionic constraints that is the first class constraint and that can be interpreted as the generator of the
$\kappa-$symmetry. Since it is the first class constraint it can be fixed
by imposing one of the fermionic variables to be equal to zero and we return to the previous case.

Finally we perform Hamiltonian analysis of superparticle in Newton-Cartan  background. This action was found in \cite{Bergshoeff:2015uaa} up to terms quadratic in fermions. We again determine Hamiltonian constraint which however
has much more complicated form  due to the presence of fermions . We also determine two sets of primary fermionic constraints. As the next step we study the requirement of the preservation of these constraints during the time evolution of the system. It turns out that this is rather non-trivial and complicated task in the full generality and hence we restrict ourselves to the simpler case of the background with the flat spatial sections. In this case we
show that the Hamiltonian constraint is the first class constraint with
vanishing Poisson brackets with fermionic constraints. We also identify linear combination of the fermionic constraints which is the first class constraint and that can be interpreted as the generator of $\kappa-$symmetry.

The extension of this paper is obvious. It would be very interesting to analyze constraint structure of superparticle in Newton-Cartan background in the full generality. Explicitly, we should analyze the time evolutions of all constraints and determine conditions on the background fields with analogy with the case of relativistic superparticle as was studied in \cite{Shapiro:1986xp}. We hope to return to this problem in future.

The organization of this paper is as follows. In the next section
(\ref{second}) we perform Hamiltonian analysis of particle in
Newton-Cartan background. Then in section (\ref{third}) we
generalize this analysis to the case of particle in
Newton-Cartan-Hooke background. In section (\ref{fourth}) we perform
Hamiltonian analysis of Galilean superparticle. Finally in section
(\ref{fifth}) we analyze Non-Relativistic Superparticle in
Cartan-Newton Background.
\section{Hamiltonian of Newton-Cartan Particle}\label{second}
By Newton-Cartan Particle we mean a particle that moves in
Newton-Cartan background with an action invariant under general
coordinate transformations. In order to describe Newton-Cartan
background we need a temporal vielbein $\tau_\mu$ and spatial
vielbein $e_\mu^{ \ a} \  , \mu=0,\dots,d-1 \ , a=1,\dots,d-1$. We
also need a central charge gauge field $m_\mu$. Then the action for
particle in Newton-Cartan background   has the form
\cite{Bergshoeff:2015uaa}
\begin{equation}\label{actnonBerg}
S=\frac{m}{2}\int d\lambda
\left[\frac{\dot{x}^\mu e_\mu^{ \ a}\dot{x}^\nu e_\nu^{ \ b}\delta_{ab}}
{\dot{x}^\rho \tau_\rho}-2m_\mu\dot{x}^\mu\right] \ ,
\end{equation}
where $\lambda$ is a parameter that labels the world-line of the particle and $\dot{x}^\mu=\frac{dx^\mu}{d\lambda}$.

 From (\ref{actnonBerg})  we derive following
conjugate momenta
\begin{equation}\label{momenta}
p_\mu= m\frac{e_\mu^{ \ a}\dot{x}^\nu e_\nu^{ \ b}\delta_{ab}}{
    \dot{x}^\rho \tau_\rho}-mm_\mu
-\frac{m}{2}\frac{\dot{x}^\rho e_\rho^{ \ a}\dot{x}^\nu
    e_{\nu}^{ \ b}\delta_{ab}}{(\dot{x}^\rho \tau_\rho)^2}\tau_\mu \ .
\end{equation}
Using (\ref{momenta}) it is easy to see that the bare Hamiltonian is equal to
zero
\begin{equation}
H_B=p_\mu \dot{x}^\mu-L=0
\end{equation}
as we could expect since the action (\ref{actnonBerg}) is manifestly invariant under world-line reparameterization
\begin{equation}
\lambda'=\lambda'(\lambda) \ , \quad  x'^\mu (\lambda')=x^\mu(\lambda) \ .
\end{equation}
The fact that the theory is invariant under world-line diffeomorphism suggests that there should exist corresponding Hamiltonian constraint. In order to find it we use following
 relations
\begin{equation}
\tau^\mu\tau_\mu=1 \ , \quad \tau^\mu e_\mu^{ \ a}=0
\end{equation}
in (\ref{momenta}) and
we obtain
\begin{eqnarray}
\tau^\mu (p_\mu+mm_\mu)&=&-\frac{m}{2}
\frac{\dot{x}^\mu e_\mu^{ \ a}\dot{x}^\nu
    e_{\nu}^{ \ b}\delta_{ab}}{(\dot{x}^\rho \tau_\rho)^2} \ , \nonumber \\
e^\mu_{ \ a}(p_\mu+mm_\mu)&=&
m\frac{\delta_{ab}e_\nu^{ \ b}\dot{x}^\nu}{\tau_\rho \dot{x}^\rho} \ .
\nonumber \\
\end{eqnarray}
If we combine these results together we obtain following primary constraint
\begin{eqnarray}\label{Hamconorg}
\mH_\lambda=
e^\mu_{ \ a}(p_\mu+mm_\mu)e^\nu_{ \ b}(p_\nu+mm_\nu)\delta^{ab}+2m \tau^\mu(p_\mu+mm_\mu) \approx 0
\nonumber \\
\end{eqnarray}
which is desired result. Then extended Hamiltonian has the form
\begin{equation}
H_T=\lambda^\lambda \mH_\lambda \ ,
\end{equation}
where $\lambda^\lambda$ is a Lagrange multiplier corresponding to the first class
constraint $\mH_\lambda$. The equations of motion for $x^\mu,p_\mu$ have the form
\begin{eqnarray}
\dot{x}^\mu&=&\pb{x^\mu,H_T}=
2\lambda^\lambda \left( h^{\mu\nu}(p_\nu+mm_\nu)+2m\tau^\mu \right)\ ,  \nonumber \\
\dot{p}_\mu&=&\pb{p_\mu,H_T}=
\lambda^\tau\left(\partial_\mu h^{\rho\sigma}(p_\rho+mm_\rho)(p_\sigma+mm_\sigma)
-2m\partial_\mu m_\rho h^{\rho\sigma}(p_\sigma+mm_\sigma)\right.-\nonumber \\
&-&\left.
2m\partial_\mu \tau^\rho (p_\rho+mm_\rho)-2m^2\tau^\rho \partial_\mu
m_\rho\right)  \ , \nonumber \\
\end{eqnarray}
where
\begin{equation}
h^{\mu\nu}=e^\mu_{ \ a}e^\nu_{ \ b}\delta^{ab} \ .
\end{equation}
As the next step  we would like to present an alternative way how to derive  an action for Newton-Cartan particle. We start with an action for  relativistic particle
in general background
\begin{equation}\label{actrel}
S=-M\int d\lambda( \sqrt{-\eta_{AB}E_\mu^{ \ A}E_\nu^{ \ B}\dot{x}^\mu
    \dot{x^\nu}}-M_\mu\dot{x}^\mu) \ ,
\end{equation}
where $E_\mu^{ \ A}$ is vielbein where $A=0,\dots,d-1$ and where exists inverse vielbein $E^\mu_{ \ A}$ defined as
\begin{equation}
E_\mu^{ \ A}E^\nu_{ \ B}=\delta_B^A \ ,  \quad
E_\mu^{ \ A}E^\nu_{ \ A}=\delta_\mu^\nu \ .
\end{equation}
Now we find corresponding canonical form of the action (\ref{actrel}). To do this we derive conjugate momenta from (\ref{actrel})
\begin{equation}
p_\mu=M\frac{\eta_{AB}E_\mu^{ \ A}E_\nu^{ \ B}\dot{x}^\nu}
{\sqrt{-\eta_{AB}E_\mu^{ \ A}E_\nu^{ \ B}\dot{x}^\mu \dot{x}^\nu}}+MM_\mu
\end{equation}
that implies following primary constraint
\begin{eqnarray}\label{constrel}
\mH_\lambda\equiv   (p_\mu-MM_\mu)E^\mu_{ \ C}\eta^{CA}E^\nu_{ \ A}
(p_\nu-MM_\nu)+M^2\approx  0  \ .
\nonumber \\
\end{eqnarray}
Now we are ready to find Hamiltonian constraint for Newton-Cartan particle
using following form of the vielbein $E_\mu^{ \ A}$ that was introduced
in \cite{Bergshoeff:2015uaa}
\begin{equation}\label{bergans}
E_\mu^{ \ 0}=\omega \tau_\mu+\frac{1}{2\omega}m_\mu \ , \quad  E_\mu^{ \ a}=
e_\mu^{ \ a} \ ,
\end{equation}
where $a=1,\dots,d-1$ and where $\omega$ is a free parameter and we  take  $\omega\rightarrow \infty$
in the end.
An inverse vielbein to (\ref{bergans}) has the form
\begin{eqnarray}\label{inbergans}
E^\mu_{ \ a}=e^\mu_{ \ a} \  , \quad  E^\mu_{ \ 0}=\frac{1}{\omega}\tau^\mu \ ,
\nonumber \\
\end{eqnarray}
where $e^\mu_{ \ a},\tau^\mu$ are defined as
\cite{Bergshoeff:2015uaa}
\begin{equation}\label{etau}
e^\mu_{ \ a}e_\mu^{ \ b}=\delta^a_b \ , \quad  \tau^\mu\tau_\mu=1 \ , \quad
\tau^\mu e_\mu^{ \ a}=\tau_\mu e^\mu_{ \ a}=0  \ , \quad
e^\rho_{ \ a}e_\mu^{ \ a}=\delta^\rho_\mu-\tau_\mu \tau^\rho \ .
\end{equation}
Note that (\ref{inbergans}) contain term linear in $\frac{1}{\omega}$. In principle it contains infinite number of terms of corrections $\frac{1}{\omega^n}$ that will give vanishing contributions in the limit $\omega\rightarrow \infty$.
%
Finally the gauge field has the form   \cite{Bergshoeff:2015uaa}
\begin{equation}\label{gaugeberg}
M_\mu=\omega \tau_\mu-\frac{1}{2\omega}m_\mu \ .
\end{equation}
Inserting (\ref{inbergans}) together with (\ref{gaugeberg}) into
(\ref{constrel}) and rescaling $M$ as $M=\omega m$ we obtain
\begin{eqnarray}
\mH_\lambda
&=&-\frac{1}{\omega^2}p_\mu \tau^\mu p_\nu \tau^\nu+p_\mu e^\mu_{ \ a}\delta^{ab}e^\nu_{ \ b}p_\nu+2mp_\mu \tau^\mu +m p_\mu h^{\mu\nu}m_\nu-\frac{1}{\omega^2}
mp_\mu\tau^\mu \tau^\nu m_\nu-\nonumber \\
&-&m^2\omega^2+m^2m_\mu \tau^\mu-\frac{m^2}{4\omega^2}(m_\mu \tau^\mu)^2
+\frac{m^2}{4}m_\mu e^\mu_{ \ a}\delta^{ab}e^\nu_{ \ b}m_b+m^2\omega^2\approx 0
\nonumber \\
\end{eqnarray}
that in the limit $\omega\rightarrow \infty $ simplifies considerably
\begin{equation}\label{Hamconsalt}
\mH_\lambda=2m(p_\mu+\frac{m}{2}m_\mu)\tau^\mu+(p_\mu+\frac{m}{2}m_\mu)
h^{\mu\nu}(p_\nu+\frac{m}{2}m_\nu)\approx 0 \ .
\end{equation}
After trivial redefinition $\frac{m_\mu}{2}\rightarrow
m_\mu$ the Hamiltonian constrain (\ref{Hamconsalt}) coincides with the
constraint (\ref{Hamconorg}). Using (\ref{Hamconsalt}) we find the canonical
form of the Newton-Cartan particle action
\begin{equation}\label{Sactcan}
S=\int d\lambda(p_\mu \dot{x}^\mu-v^\lambda \mH_\lambda) \ .
\end{equation}
It is instructive to find corresponding Lagrangian. Using (\ref{Sactcan})
we obtain
\begin{equation}\label{dotxcan}
\dot{x}^\mu=\pb{x^\mu,H_T}=2v^\lambda (m\tau^\mu+h^{\mu\nu}(p_\nu+mm_\nu))
\end{equation}
so that
\begin{equation}\label{Ldotx}
L=p_\mu\dot{x}^\mu-H_T=v^\lambda(p_\mu+mm_\mu)h^{\mu\nu}(p_\nu+mm_\nu)-
mm_\mu \dot{x}^\mu \ .
\end{equation}
Of course, this is not final form of the Lagrangian since it has to be function
of $x^\mu$ and $\dot{x}^\mu$ instead of $p^\mu$. In order to find this form we
 multiply (\ref{dotxcan}) with
 $\tau_\mu$ and use  (\ref{etau}) so that we can express $v^\lambda$ as
\begin{equation}\label{vlambda}
v^\lambda=\frac{1}{2m}\tau_\mu \dot{x}^\mu  \ .
\end{equation}
Further, if we multiply  (\ref{dotxcan}) with $h_{\mu\nu}$ we obtain
\begin{equation}
\frac{1}{2}h_{\mu\nu}\dot{x}^\nu=v^\lambda (p_\mu+mm_\mu)-v^\lambda \tau_\mu
\tau^\rho (p_\rho+mm_\rho)
\end{equation}
Inserting this result into (\ref{Ldotx}) and using again (\ref{etau})
together with (\ref{vlambda})
we finally obtain
\begin{equation}
L=\frac{m}{2}\frac{\dot{x}^\mu h_{\mu\nu}\dot{x^\nu}}{\dot{x}^\rho \tau_\rho}-mm_\mu\dot{x}^\mu
\end{equation}
that coincides with the Lagrangian defined in (\ref{actnonBerg}).
%
 \section{The Newton-Cartan-Hooke Particle}\label{third}
 In this section we perform Hamiltonian analysis of the Newton-Cartan-Hooke
 particle which is cosmological extension of the Newton-Cartan particle
 whose action was derived  in
\cite{Bergshoeff:2014gja}
 \begin{equation}\label{NCHaction}
 S=\int d\lambda \frac{m}{2}
 \left[\frac{\dot{x}^\mu e_\mu^{ \ a}\dot{x}^\nu
    e_{\nu}^{ \ b}\delta_{ab}}
 {\dot{x}^\rho \tau_\rho}-2m_\mu \dot{x}^\mu
 -\dot{x}^\rho \tau_\rho \frac{x^\mu e_\mu^{ \ a}
 x^\nu e_\nu^{ \ b}\delta_{ab}}{R^2}\right] \ ,
 \end{equation}
 where the $AdS$ radius $R^2$ is related to the cosmological constant
 $\Lambda, \Lambda<0$ as $R^2=-\frac{1}{\Lambda}$.
From
 (\ref{NCHaction}) we find conjugate momenta
 \begin{equation}
 p_\mu=m\frac{e_\mu^{ \ a}\dot{x}^\nu e_\nu^{ \ b}\delta_{ab}}
 {\dot{x}^\rho \tau_\rho}-\frac{m}{2}
 \frac{\dot{x}^\rho e_\rho^{ \ a}\dot{x}^\nu
    e_{\nu}^{ \ b}\delta_{ab}}
 {(\dot{x}^\rho \tau_\rho)^2}\tau_\mu-mm_\mu-\frac{m}{2R^2}\tau_\mu
 x^\rho e_\rho^{ \ a}x^\nu e_\nu^{ \ b}\delta_{ab} \ .
 \end{equation}
 Following analysis performed in previous section we determine various
 projectors of the combination $p_\mu+mm_\mu$
 \begin{eqnarray}
\tau^\mu (p_\mu+mm_\mu)&=&-\frac{m}{2}
 \frac{\dot{x}^\rho e_\rho^{ \ a}\dot{x}^\nu e_\nu^{ \ b}\delta_{ab}}{
 (\dot{x}^\rho \tau_\rho)^2}-\frac{m}{2R^2}x^\rho e_\rho^{ \ a}x^\nu e_\nu^{ \ b}
\delta_{ab} \ , \nonumber \\
e^\mu_{ \ a}(p_\mu+mm_\mu)&=&m\frac{\dot{x}^\nu e_\nu^{ \ b}\delta_{ba}}{\dot{x}^\rho
\tau_\rho} \ .
\nonumber \\
\end{eqnarray}
If we combine these two relations we obtain following primary
Hamiltonian constraint for the Newton-Cartan-Hooke particle
 \begin{equation}
 \mH_\lambda=2m\tau^\mu (p_\mu+mm_\mu)+(p_\mu+mm_\mu)h^{\mu\nu}
 (p_\nu+mm_\nu)+\frac{m^2}{R^2}h_{\mu\nu}x^\mu x^\nu \approx 0 \ .
 \end{equation}
  \section{Supersymmetric Generalization }\label{fourth}
 In this section we proceed to the Hamiltonian analysis of
 non-relativistic superparticles whose actions were derived in
 \cite{Bergshoeff:2014gja} and in \cite{Bergshoeff:2015uaa}.
 We begin with the simplest case which is Galilean Superparticle.
 We restrict ourselves to the case of three dimensions as in
 \cite{Bergshoeff:2014gja}.
 \subsection{The Galilean Superparticle}
 The action for Galilean Superparticle has the form
  \footnote{Note that we use Majorana representation where
    all gamma matrices are real $\gamma^\mu=(i\sigma_2,\sigma_1,\sigma_3)$, or explicitly
    $\gamma^0=\left(\begin{array}{cc}
    0 & 1 \\
    -1 & 0 \\ \end{array}\right) \ ,
    \gamma^1=\left(\begin{array}{cc}
    0 & 1 \\
    1 & 0 \\ \end{array}\right) \ , \gamma^2=
    \left(\begin{array}{cc}
    1 & 0 \\
    0 & -1 \\ \end{array}\right)$ that obey the standard relation
    $\gamma^a\gamma^b+\gamma^b\gamma^a=2\eta^{ab}\mathbf{I} \ ,
    \eta^{ab}=\mathrm{diag}(-1,1,1), \mathbf{I}=\left(\begin{array}{cc}
    1 & 0 \\
    0 & 1 \\ \end{array}\right)$.
    Then we presume that
    $\theta_-$ is real Majorana spinor so that
    \begin{equation}
    \bar{\theta}_-=\theta^T_-\gamma^0 \ .
    \end{equation}}
  \begin{equation}\label{Sgall}
 S=\int d\lambda \frac{m}{2}
 \left[\frac{\dot{x}^i\dot{x}_i}{\dot{t}}-\bar{\theta}_-
 \gamma^0 \dot{\theta}_-\right] \ .
 \end{equation}
 From (\ref{Sgall}) we determine corresponding conjugate momenta
\begin{eqnarray}
p_i&=&\frac{\partial L}{\partial \dot{x}^i}=m\frac{\dot{x}_i}{\dot{t}} \ ,  i=1,2 \ ,  \quad
p_t=\frac{\partial L}{\partial \dot{t}}=
-\frac{m}{2}\frac{\dot{x}^i\dot{x}_i}{\dot{t}^2} \ , \nonumber \\
 p_\alpha^-&=&\frac{\partial^L L}{\partial \dot{\theta}_-^{\alpha}}=
\frac{m}{2}(\bar{\theta}_-\gamma^0)_\alpha \ , \alpha=1,2  \ ,
\nonumber \\
\end{eqnarray}
 where $\partial^L$ means left-derivative. As usually we find
 that the bare Hamiltonian is zero
 \begin{equation}
 H_B=\dot{x}^ip_i+\dot{t}p_t+\dot{\theta}^\alpha_-p_\alpha^--L=0
 \end{equation}
 while the theory possesses three primary constraints
 \begin{eqnarray}
 \mG_\alpha^-&\equiv& p_\alpha^--\frac{m}{2}(\bar{\theta}_-\gamma^0)_\alpha=
 p_\alpha^-+\frac{m}{2}\theta^\beta \delta_{\beta\alpha} \approx 0 \ , \nonumber \\
 \mH_\lambda&\equiv& p_ip^i+2mp_t\approx 0 \ .
 \nonumber \\
 \end{eqnarray}
 It is important to stress that $p_\alpha^-$ and $\theta^\beta_-$ are
 Grassmann odd  variable with the graded Poisson  brackets
 \begin{equation}
 \pb{\theta^\alpha_-,p_\beta^-}=\pb{p_\beta^-,\theta^\alpha_-}=
 \delta_\alpha^\beta \ .
 \end{equation}
 Then it easy to find following
  non-zero Poisson bracket
 \begin{equation}
 \pb{\mG_\alpha^-,\mG_\beta^-}= m\delta_{\alpha\beta} \ .
 \end{equation}
 In other words  $\mG_\alpha^-$ are second class constraints. As a result we can eliminate all conjugate momenta $p_\alpha^-$ with the help of $\mG_\alpha^-=0$. Then of course we have to replace Poisson brackets between $\theta^\alpha_-,\theta^\beta_-$ with corresponding  Dirac brackets
 \begin{eqnarray}
 \pb{\theta^\alpha_-,\theta^\beta_-}_D=\pb{\theta^\alpha_-,
 \theta^\beta_-}-\pb{\theta^\alpha_-,\mG_\gamma^-}
 \frac{1}{m}\delta^{\gamma\delta}\pb{\mG_\delta^-,\theta^\beta_-}=
 -\frac{1}{m}\delta^{\alpha\beta} \ .
 \end{eqnarray}
  \subsection{$\kappa-$symmetric Galilean Superparticle}
 It is well known that the relativistic superparticle is invariant under additional fermionic symmetry called
 $\kappa-$symmetry \cite{deAzcarraga:1982dhu,Siegel:1983hh}.
 In case of strings and branes an existence of $\kappa-$
 symmetry is necessary for the correct counting of degrees of freedom. In the non-relativistic case it is known from
 the work \cite{Gomis:2004pw} that the $\kappa-$symmetry is just a St\"{u}ckelberg symmetry that acts as a shift on one of the fermionic coordinates, see also
\cite{Bergshoeff:2014gja}. Then this symmetry can be easily fixed by setting this fermionic coordinate to be equal to zero. It is instructive to see how it works in the canonical description so that we now perform Hamiltonian analysis of $\kappa-$symmetric  version of the flat Galilean superparticle action that depends on an additional fermionic coordinate $\theta_+$. The action has
the form \cite{Bergshoeff:2014gja}
 \begin{equation}\label{actkappagal}
 S=\int d\lambda\frac{m}{2}
 \left[\frac{\pi^i\pi_i}{\pi^0}-\bar{\theta}_-\gamma^0
 \dot{\theta}_-\right] \ ,
 \end{equation}
 where the line elements $\pi^0,\pi^i$ are defined as
 \begin{equation}
 \pi^0=\dot{t}+\frac{1}{4}\bar{\theta}_+\gamma^0
 \dot{\theta}_+ \ , \quad
 \pi^i=\dot{x}^i+\frac{1}{4}\bar{\theta}_-\gamma^i
 \dot{\theta}_++\frac{1}{4}\bar{\theta}_+\gamma^i
 \dot{\theta}_-
  \ .
  \end{equation}
 Now we proceed to the canonical formulation of this theory. From
 (\ref{actkappagal}) we obtain
 \begin{eqnarray}
p_i&=&\frac{\partial L}{\partial \dot{x}^i}=m\frac{\pi_i}{\pi^0} \ ,
 \quad
p_t=\frac{\partial L}{\partial \dot{t}}=-\frac{m}{2}\frac{\pi_i\pi^i}{(\pi^0)^2} \ ,
\nonumber \\
p_\alpha^-&=&\frac{\partial^L L}{\partial \dot{\theta}^\alpha_{-}}=
-\frac{m}{4}(\bar{\theta}_+\gamma^i)_\alpha\frac{\pi_i}{\pi^0}+
(\bar{\theta}_-\gamma^0)_\alpha \ , \nonumber \\
p_\alpha^+&=&\frac{\partial^L L}{\partial \dot{\theta}^\alpha_{+}}=
\frac{m}{8}(\bar{\theta}_+\gamma^0)_\alpha\frac{\pi^i\pi_i}{(\pi^0)^2}
-\frac{m}{4}\frac{\pi^i}{\pi^{0}}(\bar{\theta}_-\gamma^i)_\alpha  \ .
\nonumber \\
\end{eqnarray}
Last two equations imply two sets of primary constraints
\begin{eqnarray}
\mG_\alpha^-=
p_\alpha^-+\frac{1}{4}(\bar{\theta}_+\gamma^i)_\alpha p_i-\frac{m}{2}
(\bar{\theta}_-\gamma^0)_\alpha \approx 0 \ , \nonumber \\
\mG_\alpha^+=p_\alpha^++\frac{1}{4}(\bar{\theta}_+\gamma^0)_\alpha p_t+\frac{1}{4}
(\bar{\theta}_-\gamma^i)_\alpha p_i \approx 0 \
\nonumber \\
\end{eqnarray}
 with following non-zero Poisson brackets
\begin{eqnarray}
\pb{\mG_\alpha^-,\mG_\beta^-}=m\delta_{\alpha\beta} \ , \quad
\pb{\mG_\alpha^-,\mG_\beta^+}=\frac{1}{2}(\gamma^0\gamma^i)_{\alpha\beta}p_i \ ,
\quad
\pb{\mG_\alpha^+,\mG_\beta^+}=-\frac{1}{2}\delta_{\alpha\beta}p_t \ .
\nonumber \\
\end{eqnarray}
 Finally the theory possesses the Hamiltonian constraint in the form
 \begin{equation}
 \mH_\lambda=2mp_t+p_ip^i\approx 0
 \end{equation}
 that is clearly first class constraint
 \begin{equation}
 \pb{\mH_\lambda,\mG_\alpha^+}=\pb{\mH_\lambda,\mG_\alpha^-}=0 \ .
 \end{equation}
Again the bare Hamiltonian is equal to zero so that the extended Hamiltonian
with the primary constraints included has the form
 \begin{equation}
 H_T=\lambda^\lambda \mH_\lambda+\lambda^\alpha_+\mG_\alpha^++
 \lambda^\alpha_-\mG_\alpha^-  \ ,
 \end{equation}
 where $\lambda^\lambda,\lambda^\alpha_+,\lambda^\alpha_-$ are corresponding
 Lagrange multipliers.

 Now we have to check  consistency of all constraints. $\mH_\lambda\approx 0$
 is clearly preserved and it is the generator of world-line reparameterization. In case of the fermionic constraints  we obtain
 \begin{eqnarray}\label{eqpresG}
 \frac{d}{d\lambda} \mG_\alpha^-&=&\pb{\mG_\alpha^-,H_T}=
 \frac{1}{2} \lambda^\beta_+ (\gamma^0\gamma^i)_{\alpha\beta}p_i+
\lambda^\beta_- \delta_{\alpha\beta}m=0 \ , \nonumber \\
\frac{d}{d\lambda} \mG_\alpha^+&=&\pb{\mG_\alpha^+,H_T}=
-\lambda^\beta_+ \frac{1}{2}\delta_{\alpha\beta} p_t+\lambda^\beta_-
\frac{1}{2}(\gamma^0\gamma^ i)_{\alpha\beta}p_i=0 \ .
 \nonumber \\
 \end{eqnarray}
 From the last equation we obtain
 \begin{equation}
 \lambda^\alpha_+=\frac{1}{p_t}\delta^{\alpha\beta}(\gamma^0\gamma^ i)_{\beta\gamma}
 \lambda^\gamma_- p_i \ .
 \end{equation}
 Inserting this result
 to the first equation (\ref{eqpresG})  we find that it is obeyed since it is proportional to
 \begin{equation}
 (p_ip^i+2mp_t)=\mH_\lambda\approx 0
 \end{equation}
 so that $\lambda^\beta_-$ is not specified and it is a free parameter. Then
 it is natural to introduce following  linear combination of the fermionic constraints
\begin{equation}
\Sigma_\alpha=
\mG_\gamma^+ \delta^{\gamma\beta}(\gamma^0\gamma^i)_{\beta\alpha}\frac{1}{p_t}+\mG^-_\alpha \ .
\end{equation}
Then it is easy to see that
 \begin{eqnarray}
 \pb{\Sigma_\alpha,\mG_\beta^-}=
 \frac{1}{2p_t}
 (p_ip^i+2mp_t)\delta_{\alpha\beta}\approx 0 \ ,  \quad
  \pb{\Sigma_\alpha,\mG_\beta^+}=0
 \nonumber \\
 \end{eqnarray}
 which implies that $\pb{\Sigma_\alpha,\Sigma_\beta}\approx 0$ and hence
 $\Sigma_\alpha\approx 0$ is the first class constraint while $\mG_\alpha^-\approx 0$ is the second class constraint. Let us then  calculate the Poisson brackets between all canonical variables and $\Sigma_\alpha$
 \begin{eqnarray}\label{pbSigma}
 \pb{\Sigma_\alpha,\theta^\beta_{+}}&=&(\gamma^0\gamma^i)_\alpha^\beta\frac{p_i}{p_t} \ ,  \quad
 \pb{\Sigma_\alpha,\theta^\beta_{-}}=\delta_\alpha^\beta \ ,
 \nonumber \\
 \pb{\Sigma_\alpha,x^i}&=&-\frac{1}{4p_t}(\theta_-\gamma^i\gamma^j)_\alpha p_j-\frac{1}{4}(\bar{\theta}_+ \gamma^i)_\alpha \ , \quad
 \pb{\Sigma_\alpha,t}=\frac{1}{4}(\bar{\theta}_+\gamma^j)_\alpha
 \frac{p_j}{p_t} \ . \nonumber \\
 \end{eqnarray}
 With analogy with the Lagrangian description we would like to define
 $\kappa-$transformation
 that acts on the $\theta_+$  as a shift. For that reason we  define $\kappa-$transformation  in the following way
 \begin{equation}
 \delta_\kappa X=\frac{p_j}{2m}(\kappa \gamma^j\gamma^0)^\alpha\pb{ \Sigma_\alpha,X}\equiv \kappa^\alpha\pb{\Theta_\alpha,X} \ .
 \end{equation}
 Then using (\ref{pbSigma}) we easily find
 \begin{eqnarray}\label{kappatr}
 \delta_\kappa \theta^\alpha_+&=&\kappa^\alpha \ , \quad
 \delta_\kappa\theta^\alpha_-=\frac{p_j}{2m}(\kappa
 \gamma^j\gamma^0)^\alpha \ , \nonumber \\
 \delta_\kappa x^i&=&\frac{1}{4}
 \bar{\kappa} \gamma^i\theta_--\frac{p_j}{8m}
 \kappa \gamma^j\gamma^i\theta_+  \ , \quad
 \delta_\kappa t=-\frac{1}{4}\kappa \theta_+ \  \nonumber \\
  \end{eqnarray}
  that are correct $\kappa-$symmetry transformations. Fixing the
  $\kappa-$symmetry can be done by imposing the gauge fixing condition
  \begin{equation}
  \Omega^\alpha=\theta^\alpha_+=0
  \end{equation}
  so that
  \begin{equation}
  \pb{\Theta_\alpha,\Omega^\beta}=\delta_\alpha^\beta
  \end{equation}
  and hence we see that they are second class constraints that can be explicitly solved for $\theta_+^\alpha$ and $p_\alpha^+$. Further,
  the constraints $\mG_\alpha^-\approx 0$ are two second class constraints
  that can be solved for $p_\alpha^-$ exactly in the same way as in
  previous section.
  \section{Non-Relativistic Superpraticle in Newton-Cartan Background}\label{fifth}
  In this section we perform Hamiltonian analysis of the non-relativistic
  superparticle in Cartan-Newton Background. The action for this particle
  was derived in    \cite{Bergshoeff:2015uaa}
 and it has the form
\begin{eqnarray}
S=\frac{m}{2}\int d\lambda\left[\frac{\hpi^a\hpi^b\delta_{ab}}{\hpi^0}
-2\dot{x}^\mu (m_\mu-\bar{\theta}_-\gamma^0\psi_{\mu-})
-\bar{\theta}_-\gamma^0\hat{D}\theta_--\frac{1}{2}\dot{x}^\mu
\omega_\mu^{ \ a}\bar{\theta}_+\gamma_a \theta_-\right] \ ,
\nonumber \\
\end{eqnarray}
where we have following supersymmetric line elements
\begin{eqnarray}
\hpi^0&=&\dot{x}^\mu (\tau_\mu-\frac{1}{2}\bar{\theta}_+\gamma^0
\psi_{\mu+})+\frac{1}{4}\bar{\theta}_+\gamma^0\hat{D}\theta_+ \ ,
\nonumber \\
\hpi^a&=&\dot{x}^\mu(e_\mu^{ \ a}-\frac{1}{2}\bar{\theta}_+
\gamma^a \psi_{\mu -}-\frac{1}{2}\bar{\theta}_-\gamma^a
\psi_{\mu +})+\frac{1}{4}\bar{\theta}_+\gamma^a
\hat{D}\theta_-+\frac{1}{4}\bar{\theta}_-\gamma^a
\hat{D}\theta_++\nonumber \\
&+&\frac{1}{8}\bar{\theta}_+\gamma^a \gamma_{b0}
\theta_+ \dot{x}^\mu \omega_\mu^{ \ b} \ ,
\nonumber \\
\end{eqnarray}
 where $\hat{D}$ is covariant derivative with respect to spatial
 rotation
 \begin{equation}
 \hat{D}\theta_-=\dot{\theta}_--\frac{1}{4}\dot{x}^\mu
 \omega_{\mu}^{\ ab}\gamma_{ab}\theta_- \ .
 \end{equation}
Finally it is understood that we are interested in terms up to second order in fermions in $\hpi^a\hpi^b$. It is important to stress that spin connections
$\omega_\mu^{ \ a}, \omega_{\mu}^{ \ ab}$  depend on $e,\tau,m,\psi$. The explicit form of this dependence can  be found in \cite{Bergshoeff:2015uaa}.

 Now we can proceed to the Hamiltonian formalism. First of all we rewrite
 the line elements $\hpi^a$ and $\hpi^0$ as
 \begin{eqnarray}
 \hpi^a&=&\dot{x}^\mu \bM_\mu^{ \ a}+\frac{1}{4}\bar{\theta}_+\gamma^a \dot{\theta}_-+
 \frac{1}{4}\bar{\theta}_-\gamma^a\dot{\theta}_+ \ , \nonumber \\
 \hpi^0&=&\dot{x}^\mu \bV_\mu+\frac{1}{4}\bar{\theta}_+\gamma^0\dot{\theta}_+ \ ,
 \nonumber \\
 \end{eqnarray}
 where
 \begin{eqnarray}
 \bM_\mu^{ \ a}&=&e_\mu^{ \ a}
 -\frac{1}{2}\bar{\theta}_+
 \gamma^a \psi_{\mu -}-\frac{1}{2}\bar{\theta}_-\gamma^a
 \psi_{\mu +}-\frac{1}{16}\bar{\theta}_+\gamma^a \omega_\mu^{ \ cd}\gamma_{cd}\theta_-
 -\frac{1}{16}\bar{\theta}_-\gamma^a \omega_\mu^{\ cd}\gamma_{cd}\theta_++
 \frac{1}{8}\bar{\theta}_+\gamma^a\gamma_{b0}\theta_+\omega_\mu^{ \ b} \ ,
 \nonumber \\
 \bV_\mu&=&\tau_\mu-\frac{1}{2}\bar{\theta}_+\gamma^0\psi_{\mu+}-\frac{1}{16}
 \bar{\theta}_+\gamma^0\omega_\mu^{ \ ab}\gamma_{ab}\theta_+ \ . \nonumber \\
 \end{eqnarray}
 To proceed further we now  presume an existence of the inverse matrix $\bM^\mu_{ \ b}$ to $\bM_\mu^{ \ a}$ that obeys the relation
 \begin{equation}
 \bM^\mu_{ \ b}\bM_\mu^{ \ a}=\delta_b^a
 \ .
 \end{equation}
 In fact, since we are interested in terms up to second order in fermions
 we derive this matrix as follows. Let us denote original matrix and its inverse as
 \begin{equation}
 \bM_\mu^{ \ a}=e_\mu^{ \ a}+\bV_\mu^{ \ a}  \ ,  \quad
 \bM^\mu_{ \ b}=e^\mu_{ \ b}+\bW^\mu_{ \ b}  \ .
 \end{equation}
 Then the condition $\bM^\mu_{ \ b}\bM_\mu^{ \ a}=\delta_b^a$ implies
 \begin{eqnarray}
 \bM^\mu_{ \ b}\bM_\mu^{ \ a}=
 e^\mu_{ \ b}e_\mu^{ \ a}+e^\mu_{ \ b}\bV_\mu^{ \ a}+
 \bW^\mu_{ \ b}e_\mu^{ \ a}+O(\theta^4)=\delta_b^a
 \end{eqnarray}
 so that we obtain the condition
 \begin{equation}
 \bW^\mu_{ \ b}e_\mu^{ \ a}=-e^\mu_{ \ b}\bV_\mu^{ \ a} \ .
 \end{equation}
 In order to determine $\bW^\mu_{ \ a}$ we multiply the last equation with  $e^\nu_{ \ a}$ and we obtain
\begin{eqnarray}
\bW^\mu_{ \ b}(\delta_\mu^\nu-\tau_\mu \tau^\nu)=-e^\mu_{ \ b}\bV_\mu^{ \ a}e^\nu_{ \ a}
\nonumber \\
\end{eqnarray}
 that has solution
 \begin{equation}
 \bW^\mu_{ \ a}=-e^\nu_{ \ a}\bV_\nu^{ \ b}e^\mu_{ \ b}
 \end{equation}
 since $e^\mu_{ \ b}\tau_\mu=0$.
  Using this notation we determine corresponding conjugate momenta
 \begin{eqnarray}\label{pmupalpha}
 p_\mu&=&m\frac{\bM_\mu^{ \ a}\delta_{ab}\hpi^b}{\hpi^0}
 -\frac{m}{2}\frac{\hpi^a \hpi^b \delta_{ab}}{(\hpi^0)^2}\bV_\mu-\nonumber \\
 &-&mm_\mu+\frac{m}{8}\bar{\theta}_-\gamma^0
 \omega_\mu^{ \ ab}\gamma_{ab}\theta_--\frac{m}{4}\omega_\mu^{ \ a}
 \bar{\theta}_+\gamma_a \theta_- \ ,
 \nonumber \\
  p_\alpha^-&=&\frac{\partial^L L}{\partial \dot{\theta}^\alpha_-}=
 -\frac{m}{4}(\bar{\theta}_+\gamma^b)_\alpha \delta_{ba}\frac{\hpi^b}{
    \hpi^0}-\frac{m}{2}\theta^\beta_{-}\delta_{\beta\alpha} \ ,
 \nonumber \\
 p_\alpha^+&=&\frac{\partial^L L}{\partial \dot{\theta}^\alpha_+}=
 -\frac{m}{4}(\bar{\theta}_-\gamma^a )_\alpha
 \frac{\delta_{ab}\hpi^b}{\hpi^0}+\frac{m}{8}\frac{\hpi^a
    \hpi^b\delta_{ab}}{(\hpi^0)^2}(\bar{\theta}_+\gamma^0)_\alpha \ .  \nonumber \\
 \end{eqnarray}
 To proceed further we have to introduce vector $\bT^\mu$ that obeys the condition
 \begin{equation}\label{bTbM}
 \bT^\mu \bM_\mu^{ \ a}=0
 \ .
 \end{equation}
With analogy with the pure bosonic case we presume that it has the form
 \begin{equation}
 \bT^\mu=\tau^\mu+\bW^\mu \ ,
 \end{equation}
 where $\bW^\mu$ is quadratic in fermions. Then the condition
 (\ref{bTbM}) implies
 \begin{equation}
 \tau^\mu \bV_\mu^{ \ a}=-\bW^\mu e_\mu^{ \ a}
 \end{equation}
 that can be again solved as
 \begin{equation}
 \bW^\mu=-\tau^\nu \bV_\nu^{ \ a}e^\mu_{ \ a} \ .
 \end{equation}
 Finally note that generally
 \begin{equation}
 \bT^\mu \bV_\mu\neq 1 \ .
 \end{equation}
 With the help of the knowledge of $\bT^\mu \ , \bM^\mu_{ \ a}$ we
 can proceed to the derivation of corresponding Hamiltonian constraint.
 Using $\bT^\mu$ and $\bM_\mu^{ \ a}$ we find
 \begin{eqnarray}\label{bMPi}
& & \frac{m}{2}\frac{\hpi^a \hpi^b\delta_{ab}}{(\hpi^0)^2}=
 -\frac{1}{\bT^\mu \bV_\mu} \bT^\mu \Pi_\mu
 \ , \nonumber \\
& & \bM^\mu_{ \ a}\Pi_\mu
=
 m\frac{\delta_{ab}\hpi^b}{\hpi^0}
 -\frac{m}{2}\frac{\hpi^a\hpi^b\delta_{ab}}{(\hpi^0)^2}
 \bM^\mu_{ \ a}\bV_\mu \ ,  \nonumber \\
 \end{eqnarray}
 where we defined $\Pi_\mu$ as
\begin{equation}
\Pi_\mu=
p_\mu+mm_\mu+\frac{m}{4}\omega_\mu^{ \ a}
\bar{\theta}_+\gamma_a \theta_--\frac{m}{8}\bar{\theta}_-
\gamma^0\omega_\mu^{ \ ab}\gamma_{ab}\theta_- \ .
\end{equation}
Inserting the first  equation in (\ref{bMPi}) into the second one
we obtain
 \begin{eqnarray}
\left( \bM^\mu_{ \ a}-\frac{\bT^\mu}{\bT^\nu \bV_\nu}\bM^\rho_{ \ a} \bV_\rho\right)\Pi_\mu
= m\frac{\delta_{ab}\hpi^b}{\hpi^0}
 \nonumber \\
 \end{eqnarray}
 so that  we obtain following Hamiltonian constraint
 \begin{eqnarray}
 \mH_\lambda=
  ( \bM^\mu_{ \ a}\bT^\sigma \bV_\sigma-\bT^\mu\bM^\rho_{ \ a} \bV_\rho)\Pi_\mu
 \delta^{ab}( \bM^\nu_{ \ a}\bT^\sigma \bV_\sigma-\bT^\nu\bM^\sigma_{ \ a} \bV_\sigma)
 \Pi_\nu+2\bT^\rho\bV_\rho\bT^\mu \Pi_\mu \approx 0 \ . \nonumber \\
  \end{eqnarray}
  Further, using (\ref{bMPi}) in the second and third equation in
 (\ref{pmupalpha})   we also find  fermionic primary constraints
 \begin{eqnarray}
 \mG^-_\alpha&=&p^-_\alpha+\frac{m}{2}\theta^\beta_-\delta_{\beta\alpha}
  +\frac{1}{4}(\bar{\theta}_+\gamma^a)_\alpha
 (\bM^\mu_{ \ a}-\frac{\bT^\mu}{\bT^\nu \bV_\nu}\bM^\rho_{ \ a}\bV_\rho)\Pi_\mu\approx 0  \ ,
 \nonumber \\
 \mG^+_\alpha&=&p^+_\alpha +\frac{1}{4}(\bar{\theta_-}\gamma^a)_\alpha
 ( \bM^\mu_{ \ a}-\frac{\bT^\mu}{\bT^\nu \bV_\nu}\bM^\rho_{ \ a} \bV_\rho)
\Pi_\mu
+\frac{1}{4}(\bar{\theta}_+\gamma^0)_\alpha \bT^\mu \Pi_\mu  \approx 0 \ .  \nonumber \\
 \end{eqnarray}
 As the next step we should analyze the requirement of the preservation
 of all constraints during the time evolution of the system. It is a difficult
 task in the full generality due to the complicated form of these constraints.
 For that reason we restrict ourselves to the special case when some of the background fields vanish
\begin{equation}
 \psi_{\mu-}=\psi_{\mu+}=\omega_\mu^{ \ cd}=\omega_\mu^{ \ a}=0 \
 \end{equation}
 when we have
 \begin{equation}
 \bM_\mu^{ \ a}=e_\mu^{ \ a} \ , \bV_\mu=\tau_\mu \  .
 \end{equation}
 Then  all constraints simplify considerably
 \begin{eqnarray}
 \mH_\lambda&=&(p_\mu+mm_\mu)h^{\mu\nu}(p_\nu+mm_\nu)+2m \tau^\mu(p_\mu+mm_\mu)\approx 0
 \ , \nonumber \\
 \mG_\alpha^-&=&p_\alpha^-+\frac{m}{2}\theta^\beta_-\delta_{\beta\alpha}+
 \frac{1}{4}(\bar{\theta}_+
 \gamma^a)_\alpha e^\mu_{ \ a}(p_\mu+mm_\mu)\approx 0 \ ,
 \nonumber \\
 \mG_\alpha^+&=&p_\alpha^++\frac{1}{4}(\bar{\theta}_-\gamma^a)_\alpha e^\mu_{ \ a}(p_\mu+mm_\mu)
 +\frac{1}{4}(\bar{\theta}_+\gamma^0)_\alpha \tau^\mu(p_\mu+mm_\mu) \approx 0 \ .
 \nonumber \\
 \end{eqnarray}
 Now we have to calculate Poisson brackets between all constraints. We start with the following one
 \begin{eqnarray}
 \pb{\mG_\alpha^-,\mH_\lambda}&=&
 %
 \frac{1}{4}(\bar{\theta}_+\gamma^a)^\alpha
 ( (\partial_\rho e^\mu_{ \ a}+\partial_\rho e^\sigma_{ \ a}) h^{\rho\sigma}
 -e^\rho_{ \ a}\partial_\rho h^{\mu\sigma})(p_\mu+mm_\mu)(p_\sigma+mm_\sigma)+
 \nonumber \\
 &+&  \frac{m}{2}(\bar{\theta}_+\gamma^a)^\alpha (\partial_\rho e^\mu_{ \ a}\tau^\rho-e^\rho_{ \ a}
 \partial_\rho \tau^\mu)(p_\mu+mm_\mu) \ . \nonumber \\
 \end{eqnarray}
 It is important to stress that the fact that we restrict to the case $\omega_\mu^{\ ab}=
 \omega_\mu^{ \ a}=0$ implies, since they depend on spatial derivatives of $e_\mu^{ \ a},\tau^\mu$,  that we should impose the condition that $e_\mu^{ \ a},m_\mu,\tau^\mu$ do not depend on $x^\mu$ as well. Then the  equation given above  implies that    $\pb{\mG_\alpha^-,\mH_\lambda}=0$. In the same way we find that
 \begin{equation}
 \pb{\mG_\alpha^+,\mH_\lambda}=0 \
 \end{equation}
 and also
 \begin{eqnarray}
 \pb{\mG_\alpha^-,\mG_\beta^-}&=&m\delta_{\alpha\beta} \ ,
 \quad
 \pb{\mG^+_\alpha,\mG^+_\beta}=-\frac{1}{2}\delta^{\alpha\beta}\tau^\mu
 (p_\mu+mm_\mu) \ ,
 \nonumber \\
 \pb{\mG^-_\alpha,\mG^+_\beta}&=&\frac{1}{2}(\gamma^0\gamma^a)_{\alpha\beta}
 e^\mu_{ \ a}(p_\mu+mm_\mu) \ . \nonumber \\
 \end{eqnarray}
Finally we have to determine the time evolution of these constraints when we take into account that the total Hamiltonian is the sum of all primary constraints
 \begin{equation}
 H_T=\lambda^\lambda \mH_\lambda+\lambda^\alpha_- \mG_\alpha^-+\lambda^\alpha_+
 \mG_\alpha^+ \ .
 \end{equation}
 $\mH_\lambda$ is preserved automatically while the time evolution of $\mG^\alpha_-,\mG^\alpha_+$ is governed by following equations
 \begin{eqnarray}
 \dot{\mG}^-_\alpha&=&\pb{\mG_\alpha^-,H_T}=
\lambda^\beta_+ \frac{1}{2}(\gamma^0\gamma^a)_{\beta\alpha}e^\mu_{ \ a}
 (p_\mu+mm_\mu)+m\lambda^\beta_- \delta_{\beta\alpha}=0  \ ,
 \nonumber \\
 \dot{\mG}_\alpha^-&=&\pb{\mG_\alpha^-,H_T}
 =\frac{1}{2}\lambda^\beta_-(\gamma^0\gamma^a)_{\alpha\beta}e^\mu_{ \ a}
 (p_\mu+mm_\mu)-\frac{1}{2}\lambda^\beta_+ \delta_{\alpha\beta}
 \tau^\mu (p_\mu+mm_\mu)=0 \ .  \nonumber \\
 \end{eqnarray}
 From the last equation we express $\lambda^\beta_+$ as
 \begin{equation}
 \lambda^\beta_+=\delta^{\beta\gamma}(\gamma^0\gamma^a)_{\gamma\delta}
 \lambda^\delta_- e^\mu_{ \ a}
 (p_\mu+mm_\mu)\frac{1}{\tau^\mu(p_\mu+mm_\mu)}
 \end{equation}
 and inserting back to the first equation we obtain
 \begin{equation}
 \dot{\mG}_\alpha^-=
((p_\mu+mm_\mu)h^{\mu\nu}(p_\nu+mm_\nu)+2m\tau^\mu (p_\mu+mm_\mu))\lambda^\beta_-\delta_{\beta\alpha}=
\mH_\lambda \lambda^\beta_- \delta_{\beta\alpha}\approx 0
\end{equation}
 and we see that this equation is obeyed for all $\lambda^\beta_-$ on the constraint surface $\mH_\lambda\approx 0$.  This fact again
 implies an existence of the first class constraint in the form
 \begin{equation}
\Sigma_\alpha=\mG_\alpha^-
\tau^\mu (p_\mu+mm_\mu)
+\mG_\beta^+\delta^{\beta\gamma}(\gamma^0\gamma^a)_{\gamma
\alpha}e^\mu_{ \ a}(p_\mu+mm_\mu)
 \end{equation}
 that has following Poisson brackets with canonical variables
 \begin{eqnarray}\label{pbSigmagen}
 \pb{\Sigma_\alpha,x^\mu}
 &=&-\frac{1}{4}(\bar{\theta}_+\gamma^a)_\alpha
 e^\mu_{ \ a}\tau^\nu(p_\nu+mm_\nu)+\nonumber \\
 &+&\frac{1}{4}\tau^\mu(\bar{\theta}_+\gamma^b)_\alpha
 e^\nu_{ \ b}(p_\nu+mm_\nu)-\frac{1}{4}(\theta_-\gamma^a\gamma^b)_\alpha e^\mu_{ \ a}
 e^\nu_{ \ b}(p_\nu+mm_\nu) \ , \nonumber \\
 \pb{\Sigma_\alpha,\theta_+^\beta}&=&(\gamma^0\gamma^a)_{\alpha\omega}\delta^{\omega\beta}
 e^\mu_{ \ a}(p_\mu+mm_\mu) \ , \nonumber \\
 \pb{\Sigma_\alpha,\theta^\beta_{-}}&=&\delta_\alpha^{\beta}\tau^\mu (p_\mu+mm_\mu) \ .
 \nonumber \\
 \end{eqnarray}
 Now we define $\kappa$ variation as
 \begin{equation}
 \delta_\kappa X=
 -\frac{1}{2m\tau^\mu(p_\mu+mm_\mu)}(\bar{\kappa}\gamma^a)^\alpha e^\mu_{ \ a}(p_\mu+mm_\mu)
  \pb{\Sigma_\alpha,X} \ .
 \end{equation}
Then with the help of   (\ref{pbSigmagen}) we obtain following
transformation rules
 \begin{eqnarray}\label{kappagen}
 \delta_\kappa \theta^\alpha_+&=&
 \kappa^\alpha \ ,  \quad
 \delta_\kappa \theta_-^\alpha=-(\bar{\kappa}\gamma^a)^\alpha e^\mu_{ \ a}(p_\mu+mm_\mu) \ ,
 \nonumber \\
 \delta_\kappa x^\mu&=&
\frac{1}{8m}(\kappa \gamma^a\gamma^b\theta_+)e^\nu_{ \ a}(p_\nu+mm_\nu)e^\mu_{ \ b}-\frac{1}{4}\tau^\mu \kappa \theta_+
-\frac{1}{4}(\bar{\kappa}\gamma^a\theta_-)e^\mu_{ \ a}
\nonumber \\
 \end{eqnarray}
 that are covariant form of   $\kappa-$transformations
 (\ref{kappatr}). To see this explicitly note that in the
 flat non-relativistic space-time we can choose $\tau^0=1 \ , \tau^i=0,
 e_\mu^{ a}=\delta_i^a$ and the $\kappa-$transformations
 given in (\ref{kappagen}) have explicit form
   \begin{eqnarray}
 \delta_\kappa \theta_+^\alpha&=&
 \kappa^\alpha \ ,  \quad
 \delta_\kappa \theta^\alpha_-=-(\bar{\kappa}\gamma^i)^\alpha (p_i+mm_i) \ ,
 \nonumber \\
 \delta_\kappa x^i&=&
 \frac{1}{8m}(\kappa \gamma^j\gamma^i)(p_j+mm_j)
 -\frac{1}{4}(\bar{\kappa}\gamma^i\theta_-) \ ,  \quad
 \delta_\kappa t=-\frac{1}{4}\kappa \theta_+
 \nonumber \\
 \end{eqnarray}
 that coincide with the transformations (\ref{kappatr}) which is a  nice
 consistency check.

 In this section we performed Hamiltonian analysis of the non-relativistic
 superparticle in Newton-Cartan background. We derived general form of the Hamiltonian and fermionic constraints. Then we studied their properties
 for special configurations of background fields. It would be extremely interesting to analyze this theory for general background. We expect that the
 requirement of the existence of three first class constraints, where one is Hamiltonian constraint while remaining two constraints correspond to generators of $\kappa-$symmetry, will impose some restriction on the background fields.
 We hope to return to this analysis in near future.

\acknowledgments{This  work  was
    supported by the Grant Agency of the Czech Republic under the grant
    P201/12/G028. }


\end{document}